# Optically Activated Superconductivity in $MgB_2$ via Electroluminescent GaP Inhomogeneous Phase

Yao Qi, Duo Chen, Qingyu Hai, Xiaoyan Li, Xiaopeng Zhao*

Smart Materials Laboratory, Department of Applied Physics,

Northwestern Polytechnical University, Xi'an 710129, China;

qiyao@mail.nwpu.edu.cn (Y.Q.);

chenduo@mail.nwpu.edu.cn (D.C.);

haiqingyu@mail.nwpu.edu.cn (Q.H.);

lixiaoyan0521@mail.nwpu.edu.cn (X.L.);

* Correspondence: xpzhao@nwpu.edu.cn (X.Z.);

Yao Qi and Duo Chen contributed equally to this work.

# Abstract

Experimental results demonstrate a viable strategy for tuning the superconducting properties of $MgB_2$ through the incorporation of an electroluminescent inhomogeneous phase, revealing an interfacial light-phonon-electron synergistic mechanism that enhances superconductivity in conventional phonon-mediated systems. By introducing GaP electroluminescent inhomogeneous phases into $MgB_2$ and activating their emission in situ through the application of a bias current during measurements, it is experimentally observed that the localized optical field and electromagnetic near field generated at the interface can effectively couple with the $E_{2g}$ phonon mode of the Mg-B layers, thereby significantly enhancing the electron-phonon interaction. As the emission intensity of the inhomogeneous phase increases, the interface light-field-driven mechanism markedly enhances the electron-phonon coupling constant $\lambda$ and leads to a gradual increase in the superconducting transition temperature $T_c$ (with a maximum enhancement of approximately 1.4 K), enabling a tunable enhancement of the superconducting pairing channel in $MgB_2$ without altering its primary chemical composition. In addition, the nanoscale dispersed distribution of the GaP inhomogeneous phase is expected to induce fine-scale defects that act as effective pinning centers and promote densification, resulting in an increase of the critical current density by approximately 69% at 20 K in the self-field and an enhancement of $H_{irr}$ by about 31.5%. These results indicate that the electroluminescent inhomogeneous phase can synergistically enhance the superconducting performance of $MgB_2$ through two mechanisms: "in situ near-field-enhanced pairing" and "structural pinning-assisted flux optimization," thereby providing a new design strategy for constructing superconducting material systems that can be activated by internal optical fields.

# 1.Introduction

Layered structural systems play a central role in modern superconductivity research. From moiré-band engineering in magic-angle graphene systems[1, 2] to the layered coordination geometries in nickel-based and iron-based superconductors[3-6], and to the alternating Mg and B layers forming the two-dimensional framework of $MgB_2$, the electronic properties and quantum behavior of these materials are strongly governed by the band structures and vibrational modes dictated by their layered crystal architectures. In the $MgB_2$ system in particular, its characteristic two-band superconductivity and strong electron-phonon coupling originate directly from the unique soft-mode behavior of the $E_{2g}$ phonon associated with the B-B σ-bonding network[7, 8].

Among various superconducting systems, $MgB_2$, as a representative electron-phonon-coupled superconductor, has attracted extensive attention since its discovery because of its high critical temperature of 39 K, weak anisotropy, low cost, and excellent processability, which together make it highly suitable for superconducting magnets, power devices, and quantum applications[9]. However, the superconducting properties of $MgB_2$ are highly sensitive to the electron-phonon coupling strength and grain connectivity. Conventional approaches such as chemical doping, nanoparticle inclusion, and microstructural engineering may improve certain performance metrics, yet they commonly suffer from issues including $T_c$ suppression caused by lattice distortion, chemical instability, interfacial side reactions, and enhanced grain-boundary scattering[10-15]. Therefore, enhancing electron-phonon coupling and improving superconducting performance without perturbing the intrinsic lattice structure remains a critical scientific challenge in $MgB_2$ research.

In recent years, interfacial electromagnetic excitation, plasmonic near-field effects, and light-phonon cooperative interactions have demonstrated the capability to modulate electronic structures and coupling mechanisms in condensed-matter systems, offering a new perspective beyond conventional superconductivity-control pathways[16-21]. Under optical near-field excitation, lattice vibrations can be dynamically tuned, and electron-phonon interactions may be enhanced by modifying the local electromagnetic environment[22-25]. However, such light-field-based modulation typically relies on pulsed light sources or complex external-field conditions, making it difficult to implement directly in bulk superconductors[26]. The ability to create internally embedded, stable, and continuously excitable luminescent interfaces capable of generating localized optical fields would thus open new material systems and physical mechanisms for superconductivity regulation.

Guided by the design principles of intelligent metamaterials, our previous work proposed a strategy termed "electroluminescent inhomogeneous phase-enhanced superconductivity." By introducing electroluminescent inhomogeneous phases into $MgB_2$ or cuprate superconductors such as Bi(Pb)SrCaCuO and applying bias currents during electrical transport measurements, we demonstrated that the resulting interfacial electroluminescence and localized electromagnetic near-fields can construct smart meta-superconductors (SMSCs)[27-36]. We employed several types of luminescent inhomogeneous phases, including $Y_2O_3:Eu^{3+}$ + Ag particles and GaN p-n junctions, and experimentally verified that current-induced interfacial emission can enhance superconducting performance across different superconducting systems. We further proposed a physical model in which interfacial plasmons and their evanescent electromagnetic fields strengthen electron pairing[30]. In our previous work, we systematically investigated the effects of the inhomogeneous phase particle size and addition concentration on the superconducting properties, identifying 0.5 wt.% as the optimal addition level that balances the interfacial excitation effect with the connectivity of $MgB_2$[32, 33]. Therefore, in the present study, the content of the GaP inhomogeneous phase was fixed at 0.5 wt.% to minimize the interference arising from compositional or concentration variables, allowing the investigation to focus specifically on the influence of "tunable electroluminescence intensity" on the superconducting performance. However, the electroluminescent inhomogeneous phases used thus far exhibit limitations in emission intensity, directionality, and crystalline symmetry. Notably, the systematic influence of increased emission intensity on superconducting behavior has yet to be experimentally elucidated.

Building upon these prior studies, the present work introduces isotropic GaP electroluminescent nanoparticles whose emission intensity can be tuned through periodic structural modulation to systematically investigate how variations in luminescence intensity affect the structure, vibrational dynamics, electrical transport behavior, and magnetic hysteresis of $MgB_2$. We reveal an interfacial light-field-driven mechanism that enhances electron-phonon coupling. Our results demonstrate that, with increasing emission intensity, the interfacial optical field under bias excitation significantly softens the $E_{2g}$ phonon mode and increases the electron-phonon coupling constant $\lambda$, thereby leading to a steady enhancement of $T_c$. Meanwhile, the nanoscale dispersed distribution of the GaP inhomogeneous phase and the associated fine-scale defects/interfaces it introduces can act as effective flux pinning centers and promote densification, thereby improving $J_c$ and the pinning force and enabling the synergistic optimization of key superconducting parameters. These findings establish electroluminescent inhomogeneous phases as an effective platform for light-superconductor coupling and offer a new physical route toward designing superconducting materials with tunable phononic and electronic dynamics.

## 2. Results and Discussion

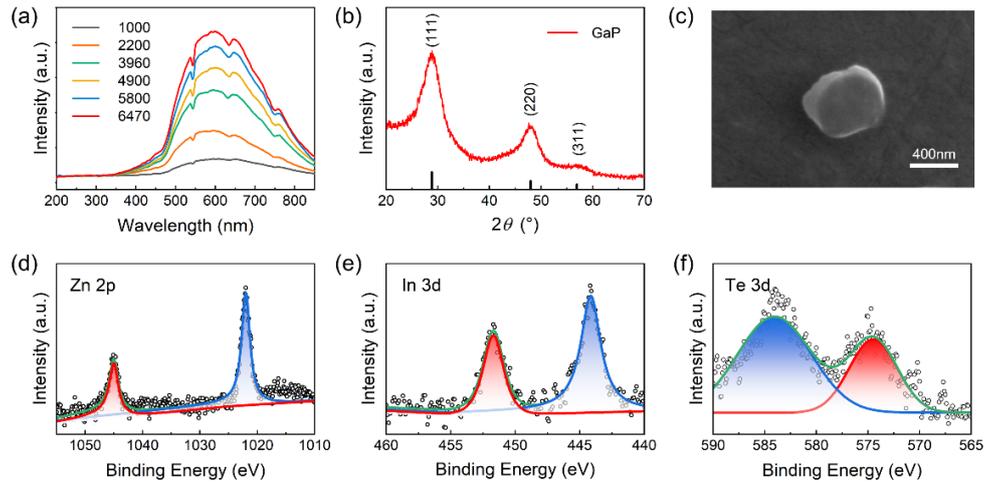

**Figure 1.**

**(a)** Electroluminescence spectrum of the GaP electroluminescent inhomogeneous phases. Strong broadband emission is retained even after high-temperature treatment, demonstrating excellent optical and structural stability.

**(b)** XRD pattern of the GaP quantum dots, showing the characteristic zinc-blende diffraction peaks at (111), (220), and (311) with no detectable impurity phases, indicating homogeneous solid-solution incorporation of the dopant elements.

**(c)** SEM image of the GaP electroluminescent inhomogeneous phase after high-temperature treatment.

**(d-f)** XPS spectra of the GaP quantum dots. Distinct characteristic peaks are observed in the Zn 2p, In 3d, and Te 3d regions, confirming the successful incorporation of the corresponding dopant elements into the GaP lattice.

Figure 1 presents the structural characteristics and optical behavior of the GaP electroluminescent inhomogeneous phase after high-temperature treatment. The resulting particles comprise two types of multilayer core-shell units, namely [GaP:Zn/GaP]$_{\times m}$/[GaInP/GaP]$_{\times n}$ and [GaP:Te/GaP]$_{\times m}$/[GaInP/GaP]$_{\times n}$, which further assemble into nanoscale composite particles during synthesis. Figure 1a presents the electroluminescence spectra of different GaP inhomogeneous phases measured under the same current-bias condition. All samples exhibit a broad emission band spanning approximately 500-750 nm, with the emission center located near 600 nm, indicating their capability to provide a stable interfacial optical field within the visible spectral range. To facilitate a quantitative correlation between the "optical field intensity" and the superconducting response in subsequent analyses, the integrated area of the emission spectra within the 500-750 nm range (after background subtraction and normalized to arbitrary units, a.u.) is defined as the EL intensity. Accordingly, the GaP batches used for composite fabrication are labeled as EL = 0, 1000, 2200, 3960, 4900, 5800, and 6600. Since the subsequent electrical transport and Raman measurements were all conducted under the same bias current, the EL intensity can be regarded as a relative representation of the near-field amplitude at the $GaP/MgB_2$ interface. Morphological characterization (Figure 1c) shows that the particles form uniform spherical aggregates with a particle size distribution centered at approximately 400 nm, exhibiting good monodispersity. This high degree of size uniformity is important for maintaining the stability of the luminescence process. The XRD pattern (Figure 1b) displays characteristic zinc-blende GaP diffraction peaks at (111), (220), and (311), with no detectable impurity phases or extraneous reflections. This indicates that the dopant elements Zn, Te, and In are successfully incorporated into the GaP host lattice to form homogeneous solid solutions. The construction of the core-shell architecture does not induce any phase transformation, reflecting favorable lattice matching. The absence of noticeable peak shifts further suggests that heterovalent doping does not introduce measurable lattice distortion, which is crucial for preserving the electronic structure of the quantum-dot units. Complementary XPS measurements (Figure 1d-f) reveal well-defined characteristic peaks in the Zn 2p, Te 3d, and In 3d regions, demonstrating that the dopant elements are incorporated into the GaP lattice in chemically stable states, consistent with the XRD analysis. Taken together, the multilayer core-shell architecture ensures high luminescence efficiency and superb thermal stability, while the favorable lattice compatibility enables this electroluminescent inhomogeneous phase to serve as a structurally stable and functionally active component for interfacial light-field coupling in $MgB_2$.

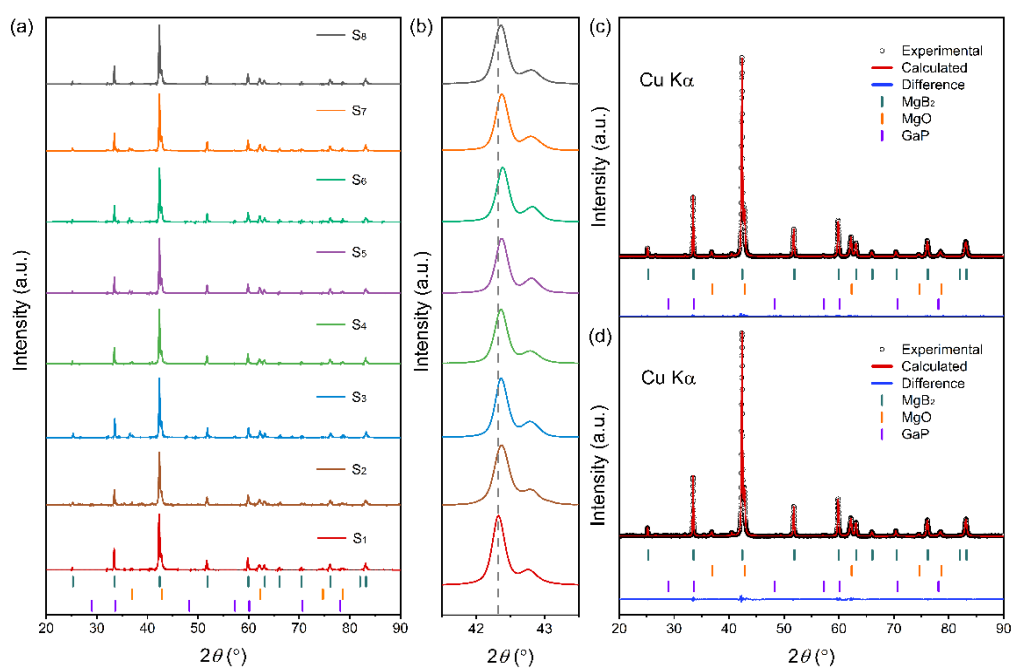

**Figure 2.**
**(a)** X-ray diffraction (XRD) patterns of the pristine $MgB_2$ sample ($S_1$) and $MgB_2$ composite samples containing 0.5 wt.% GaP electroluminescent inhomogeneous phases ($S_2$-$S_8$). The electroluminescence intensity of GaP increases from 0 to 6600 across $S_2$-$S_8$. All samples exhibit the characteristic hexagonal $MgB_2$ reflections along with minor MgO impurity peaks, while no distinct GaP diffraction signals are detected.
**(b)** Enlarged view of the $MgB_2$ (101) diffraction peak region.

**(c, d)** Rietveld-refined XRD patterns of $S_1$ and $S_8$, confirming that the incorporation of GaP does not alter the primary $MgB_2$ crystal structure. The slight variations in lattice parameters correlate with microstructural evolution and phonon dynamics within the samples.

| X (wt.%) | EL Intensity | a (Å) | c (Å) | Crystallite size (μm) | FWHM(101) | MgO(%) | $T_c$(K) |
|---|---|---|---|---|---|---|---|
| 0 | 0 | 3.0853(3) | 3.5234(1) | 0.109 | 0.214 | 10.4 | 38.2 |
| 0.5 | 0 | 3.0875(1) | 3.5265(2) | 0.119 | 0.228 | 11.5 | 37.4 |
| 0.5 | 1000 | 3.0883(1) | 3.5284(2) | 0.121 | 0.230 | 11.7 | 38.2 |
| 0.5 | 2200 | 3.0877(1) | 3.5268(1) | 0.118 | 0.227 | 12.1 | 38.5 |
| 0.5 | 3960 | 3.0871(2) | 3.5259(2) | 0.123 | 0.231 | 11.5 | 38.8 |
| 0.5 | 4900 | 3.0867(1) | 3.5263(1) | 0.117 | 0.228 | 12.0 | 39.3 |
| 0.5 | 5800 | 3.0885(1) | 3.5277(1) | 0.121 | 0.226 | 12.3 | 39.4 |
| 0.5 | 6600 | 3.0883(1) | 3.5272(1) | 0.112 | 0.225 | 11.6 | 39.6 |

**Table 1.** Rietveld refinement parameters and superconducting properties of pristine $MgB_2$ ($S_1$) and $MgB_2$ composite samples containing 0.5 wt.% GaP electroluminescent inhomogeneous phases ($S_2$-$S_8$).

Figure 2 shows the X-ray diffraction (XRD) patterns of the pristine $MgB_2$ sample ($S_1$) and the $MgB_2$ composite samples containing 0.5 wt.% GaP electroluminescent inhomogeneous phases ($S_2$-$S_8$). All samples exhibit the characteristic hexagonal $MgB_2$ diffraction features, with only minor MgO impurity peaks (approximately 10-12 wt.%) detected in addition to the main reflections, indicating that the primary crystalline phase remains stable and that the incorporation of GaP does not alter the $MgB_2$ host structure. No GaP-related diffraction peaks are observed, mainly because its concentration (0.5 wt.%) falls below the detection limit of XRD, combined with its small particle size and the partial overlap of its reflections with those of $MgB_2$. Moreover, no new diffraction peaks associated with Ga- or P-containing secondary phases are detected, confirming that GaP retains its chemical and structural stability under the sintering conditions and persists as an intact inhomogeneous phase capable of forming well-defined GaP/$MgB_2$ interfaces. It should be noted that MgO is commonly present in bulk $MgB_2$, mainly originating from the surface oxidation layer of the $MgB_2$ precursor powders and from residual $O_2/H_2O$ during the sintering process. Even when weighing and mixing are performed in a glove box, it remains difficult to completely prevent ppm-level oxygen ingress during subsequent sealing and high-temperature treatment. For the composite samples, the surfaces of GaP particles may carry small amounts of oxides or organic residues; oxygen released from these species during the 850°C heat treatment can further promote local MgO formation. Consequently, the MgO content obtained from Rietveld refinement is slightly higher in the composite samples (Table 1).

Subtle shifts in the diffraction peaks reveal slight lattice modulation of $MgB_2$ upon GaP incorporation. Rietveld refinement of the diffraction data (Table 1) shows that the a-axis lattice parameter increases marginally from 3.0853 Å to approximately 3.0883 Å, accompanied by a slight increase in the c-axis, resulting in a modest decrease in the c/a ratio. Such coexistence of in-plane expansion and mild interlayer contraction is typically associated with local strain fields introduced by nanoscale inhomogeneous phases during the sintering process, suggesting that the nanoscale dispersion of GaP within the matrix imposes a weak structural perturbation on the $MgB_2$ lattice[37]. Notably, this minor evolution in lattice parameters does not compromise the crystallinity of the samples. The full width at half maximum (FWHM) of the (101) reflection exhibits only negligible variation, and the average crystallite size remains in the range of 0.11-0.12 μm, indicating that no significant structural defects or grain-boundary degradation are introduced in this composite system. These observations confirm that the GaP inhomogeneous phase does not disrupt the $MgB_2$ host lattice or induce severe strain accumulation but instead modulates its microstructure in a mild and controlled manner.

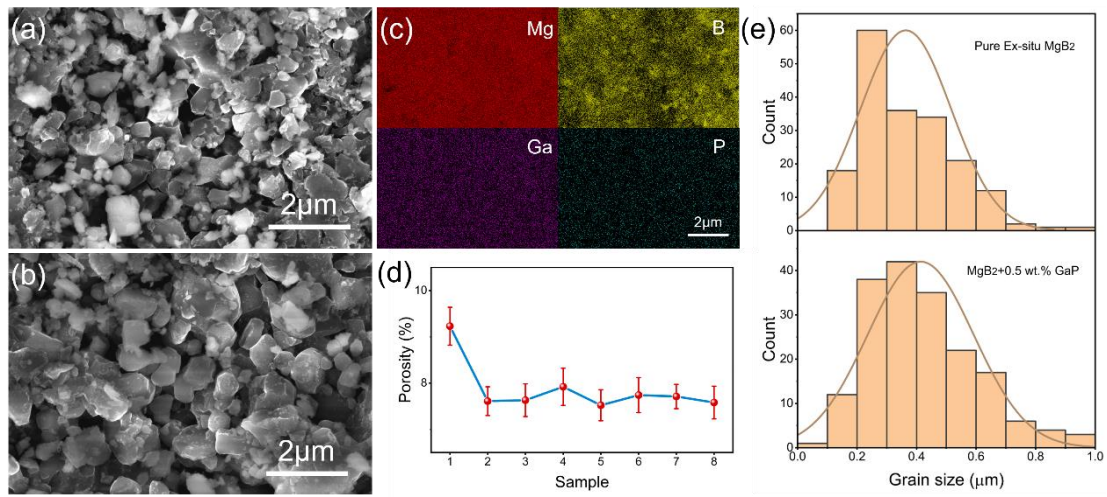

**Figure 3.**
**(a)** SEM image of the pristine $MgB_2$ sample ($S_1$), showing polycrystalline grains formed by agglomerated nanoparticles.
**(b)** SEM image of the composite sample containing 0.5 wt.% GaP (EL = 6600, $S_8$), exhibiting larger grain sizes, more rounded grain boundaries, and denser interfacial contact.
**(c)** EDS elemental mapping of the composite sample ($S_8$), showing uniformly distributed Mg/B elements and dispersed Ga/P signals.
**(d)** Porosity statistics of all samples, demonstrating an overall reduction in porosity upon introducing GaP electroluminescent inhomogeneous phases.
**(e)** Grain-size distribution histograms of the pristine sample ($S_1$) and the composite sample ($S_8$), indicating that GaP addition promotes grain growth and yields a more concentrated size distribution.

Figure 3 presents the microstructural morphology, elemental distribution, and porosity and grain-size statistics of the pristine $MgB_2$ sample and the $MgB_2$ sample incorporating 0.5 wt.% GaP electroluminescent inhomogeneous phases (EL = 6600).

The pristine MgB$_2$ sample (Figure 3a) exhibits a typical granular polycrystalline morphology in which the grains consist of agglomerated nanoscale primary particles forming a dense subgrain-network through intimate interparticle contacts. This microstructure ensures good connectivity of the matrix while still containing a limited amount of residual porosity. In contrast, the sample containing GaP inhomogeneous phases (Figure 3b) shows more fully developed grains with rounder particle morphology, tighter intergranular bonding, and noticeably larger grain sizes. These features suggest that the GaP electroluminescent particles function as heterogeneous nucleation sites for MgB$_2$ growth, becoming encapsulated and embedded within the matrix and thereby promoting grain coarsening and grain-boundary densification during sintering. Because the GaP particles are uniformly distributed, their interfacial effects act cooperatively throughout the matrix, resulting in a more stable and compact grain-boundary framework that supports the formation of continuous current-carrying pathways.

The EDS elemental mapping of the 0.5 wt.% GaP (EL = 6600) sample (S$_8$) is shown in Figure 3c. Mg and B elements are evenly distributed across the matrix, while the Ga and P signals appear as weakly dispersed features localized within nanoscale regions, with no evidence of aggregation or phase segregation. This observation further confirms that GaP inhomogeneous phases remain structurally intact and uniformly embedded at the nanoscale within the MgB$_2$ matrix without decomposition. It should be noted that EDS elemental mapping typically employs a normalized color scale to display the relative count intensity of elements; therefore, even a small amount of high-atomic-number elements (such as Ga) can exhibit relatively strong contrast. In addition, GaP tends to disperse preferentially along grain boundaries and pores. Since the two-dimensional elemental mapping covers a large grain-boundary network, relatively "widespread" Ga/P signals can be observed. Based on the nominal addition of 0.5 wt.%, the estimated volume fraction of GaP is only about ~0.3 vol.%, which is consistent with the absence of any obvious enrichment or agglomerated secondary phase. Porosity statistics for all samples (Figure 3d) reveal a marked reduction in porosity upon introducing GaP electroluminescent particles, demonstrating that the inhomogeneous phase effectively enhances the densification process. This improvement can be attributed to the dual role of GaP nanoparticles during sintering: acting as "liquid-phase bridging" sites and serving as "void fillers," both of which facilitate grain-grain contact and pore closure, thereby increasing the overall compactness of the material. Grain-size statistics (Figure 3e) further elucidate the microstructural evolution. The pristine $MgB_2$ sample exhibits a relatively broad grain-size distribution with an average size of 0.35 μm, whereas the GaP-containing sample shows a clear shift toward larger grain sizes with a narrower distribution and an increased average size of 0.45 μm. This indicates that the presence of the inhomogeneous phase moderately promotes grain growth by reducing interfacial energy at grain boundaries, which in turn helps decrease grain-boundary scattering and mitigates current-blocking effects. Combining the XRD and SEM analyses, it is evident that the GaP inhomogeneous phase, while preserving the MgB$_2$ host crystal structure, effectively improves lattice connectivity by modulating microstructural morphology, porosity, and grain development.

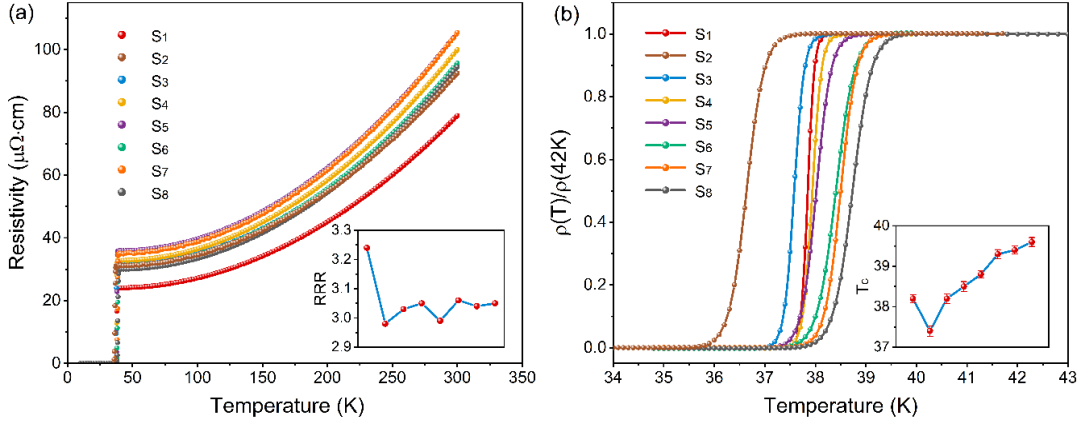

**Figure 4.**
**(a)** Temperature-dependent resistivity curves (0-300 K) of pristine MgB$_2$ and MgB$_2$ composites containing 0.5 wt.% GaP electroluminescent inhomogeneous phases. Inset: variation of the residual resistivity ratio (*RRR*) with GaP emission intensity. The composites exhibit slightly higher $\rho$(300 K) yet retain metallic conduction, with *RRR* stabilized around 3.
**(b)** Enlarged view of the superconducting transition region. Inset: $T_c$ as a function of GaP emission intensity. As the emission intensity increases, $T_c$ rises from 38.2 K to 39.6 K while the transition width remains nearly constant. Independent sample replications (batch variation < ±0.08 K) confirm that the $T_c$ enhancement originates from the electroluminescent inhomogeneous phases rather than sample-to-sample variation.

| G | EL Intensity | Doping Concentration (wt.%) | $T_c$ (K) | $\Delta T_c$ (K) | $\delta T_c$ (K) | *RRR* | $A_F$ |
|---|---|---|---|---|---|---|---|
| S$_1$ | 0 | 0 | 38.2 | 0.7 | 0 | 3.24 | 0.135 |
| S$_2$ | 0 | 0.5 | 37.4 | 1.6 | -0.8 | 2.98 | 0.104 |
| S$_3$ | 1000 | 0.5 | 38.2 | 1.3 | 0 | 3.03 | 0.115 |
| S$_4$ | 2200 | 0.5 | 38.5 | 1.2 | 0.3 | 3.05 | 0.109 |
| S$_5$ | 3960 | 0.5 | 38.8 | 1.6 | 0.6 | 2.99 | 0.104 |
| S$_6$ | 4900 | 0.5 | 39.3 | 1.8 | 1.1 | 3.06 | 0.113 |
| S$_7$ | 5800 | 0.5 | 39.4 | 1.9 | 1.2 | 3.04 | 0.103 |
| S$_8$ | 6600 | 0.5 | 39.6 | 1.8 | 1.4 | 3.05 | 0.115 |

**Table 2.** Superconducting transition temperatures and transport parameters of pristine MgB$_2$ and MgB$_2$ composites containing 0.5 wt.% GaP electroluminescent inhomogeneous phases.

Figure 4 presents the temperature-dependent resistivity curves (0-300 K) and superconducting transition characteristics of the pristine $MgB_2$ sample and the sample incorporating 0.5 wt.% GaP electroluminescent inhomogeneous phases. The pristine $MgB_2$ sample exhibits a room-temperature resistivity $\rho(300\ K)$ of approximately 78 μΩ·cm, followed by a gradual decrease upon cooling and a sharp drop to zero near 38.2 K. This metallic behavior and abrupt superconducting transition indicate high crystallinity and excellent intergrain connectivity. In comparison, the sample containing 0.5 wt.% GaP inhomogeneous phases shows a slightly higher $\rho(300\ K)$. This increase arises primarily from the interfacial potential barriers and additional scattering centers formed between the GaP nanoparticles and the $MgB_2$ matrix: after sintering, the uniformly distributed GaP particles reside preferentially at grain boundaries, and their electronic structure is not fully aligned with the metallic conduction band of $MgB_2$. As a result, carriers experience barrier scattering and local electric-field perturbations when traversing these interfaces at room temperature, leading to an increase in $\rho(300\ K)$. Nevertheless, all composite samples preserve a clear metallic $\rho$-$T$ trend, demonstrating that the macroscopic current-carrying pathways remain continuous and that GaP addition does not trigger non-metallic or localized transport behavior.

Despite the increase in $\rho(300\ K)$, the residual resistivity ratio (*RRR*) remains within a narrow range of 3.0-3.2 (Table 2), comparable to that of pristine $MgB_2$. The stability of *RRR* implies that the intrinsic intragrain scattering mechanisms and electronic transport quality are preserved, and the primary modifications occur within the grain-boundary regions. To quantitatively evaluate the effective cross-sectional area of the macroscopic current paths, the Rowell connectivity analysis was employed:

$$\Delta\rho = \rho_{(300K)} - \rho_{(40K)}$$
$$A_F = \Delta\rho_{ideal}/\Delta\rho$$

where $\Delta\rho_{ideal}$ is taken as 7.3 μΩ·cm for an ideally dense $MgB_2$[38]. With the introduction of the GaP inhomogeneous phases, $\Delta\rho$ decreases slightly, leading to a small fluctuation of $A_F$ within the range of 0.103-0.115. The limited magnitude of this variation indicates that the effective conductive cross-section of the samples remains largely stable. Combined with the densification and grain-boundary rounding observed in SEM images, the slight variation in $A_F$ is more likely to reflect the modulation of local interfacial potential barriers and grain-boundary scattering by GaP, rather than a significant degradation of the macroscopic connectivity.

The superconducting transition provides further insight into the impact of GaP inhomogeneous phases on the pairing process. The pristine $MgB_2$ sample exhibits a narrow transition width $\Delta T_c \approx 0.7$ K, with a sharp and well-defined zero-resistance transition. For the sample $S_2$ containing non-emissive GaP inhomogeneous phases, $T_{c,onset}$ is slightly suppressed relative to the pristine sample, and the transition becomes marginally broader. This behavior agrees with prior reports on inert nanostructured inhomogeneous phases, in which interfacial scattering and local strain mildly hinder the pairing process[15, 39-41]. As the GaP electroluminescence intensity increases from 1000 to 6600, $T_c$ rises steadily from 38.2 K to 39.6 K, and $T_{c,zero}$ shifts accordingly, surpassing that of the pristine sample. This simultaneous upward shift of both $T_{c,onset}$ and $T_{c,zero}$ indicates that the entire transition is elevated rather than only its high-temperature edge. Such behavior, which involves slight broadening together with a uniform upward shift, suggests a modified pairing environment. Under applied electric fields, the GaP inhomogeneous phases generate electroluminescence and surface electromagnetic polarization modes. The resulting near-field evanescent waves produce localized electromagnetic coupling zones at the GaP/MgB$_2$ interfaces, which interact with the electron-phonon system of the Mg-B layers[16, 17, 30, 33]. These interfacial near fields enhance the phase coherence of electron-phonon interactions, leading to a more uniform distribution of pairing energy and causing $T_{c,onset}$ and $T_{c,zero}$ to rise simultaneously[19]. Meanwhile, slight variations in local coupling strength among grains or regions produce mild broadening of $\Delta T_c$ without inducing significant inhomogeneity or defect scattering, indicating that the near-field effects are cooperative rather than disruptive.

Overall, the influence of GaP electroluminescent inhomogeneous phases on the electrical transport behavior exhibits multidimensional coupling effects: increases in room-temperature resistivity stem from interface scattering; the stability of $RRR$ and $A_F$ confirms that intragrain quality and macroscopic connectivity remain excellent; and the monotonic enhancement of $T_c$ reflects the strengthening of electron pairing interactions through interfacial electromagnetic near-field coupling. The localized light-field-phonon coupling channels generated by GaP under electrical excitation introduce an additional pairing enhancement mechanism, enabling effective tuning of the superconducting transition temperature without altering the intrinsic $MgB_2$ lattice.

To ensure the reliability of the observed $T_c$ enhancement, key samples were independently re-synthesized and re-measured. The reproducibility in $T_{c,onset}$ and $T_{c,zero}$ matches the primary measurements, with batch-to-batch variations within ±0.08 K, demonstrating excellent consistency. The measurement uncertainty, dominated by temperature control and contact resistance, is estimated to be ±0.07-0.10 K, which is significantly smaller than the 1.4 K systematic increase in $T_c$ induced by luminescence strengthening.

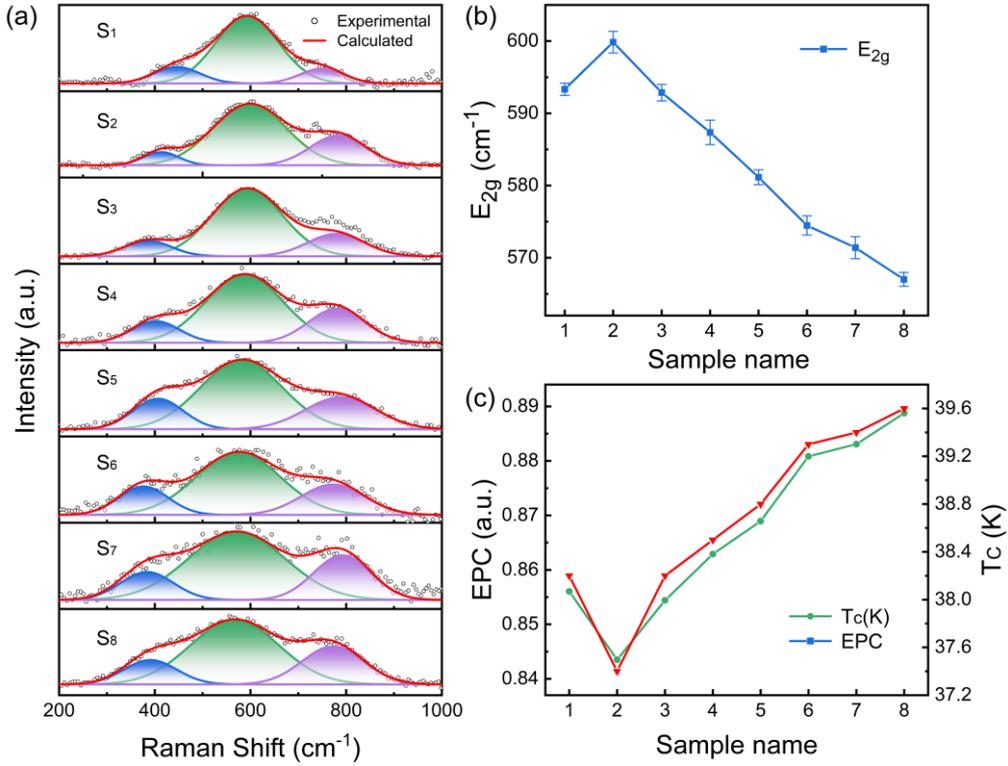

**Figure 5.**

**(a)** Raman spectra and Gaussian fitting results of pristine $MgB_2$ and $MgB_2$ composites containing 0.5 wt.% GaP electroluminescent inhomogeneous phases under a 100 mA bias current.
**(b)** Systematic evolution of the $E_{2g}$ phonon frequency with increasing GaP emission intensity, revealing progressive phonon softening.
**(c)** Electron-phonon coupling constant $\lambda$ calculated using the Allen-Dynes formula and its concurrent increase with $T_c$, demonstrating that interfacial electroluminescence-induced electromagnetic near fields enhance electron-phonon interactions in $MgB_2$.

Figure 5 shows the Raman spectra and corresponding three-peak Gaussian fitting results for the pristine $MgB_2$ sample and the $MgB_2$ composites containing 0.5 wt.% GaP electroluminescent inhomogeneous phases (EL = 1000-6600). All measurements were carried out under a 100 mA bias current to activate the GaP electroluminescent centers and generate interfacial electroluminescence and localized electromagnetic fields. The Raman spectra over the 200-1000 $cm^{-1}$ range can be accurately decomposed into three Gaussian components. The broad mid-frequency peak corresponds to the in-plane B-B bond stretching vibration of the $E_{2g}$ phonon mode, the hallmark optical mode of $MgB_2$ that is strongly linked to its electron-phonon coupling. The low-frequency peak mainly reflects acoustic-branch vibrations or defect-activated modes associated with grain boundaries, defects, and nanoparticles, whereas the high-frequency shoulder is generally attributed to second-order scattering, interface-related vibrational states, or locally disordered structures[8, 42, 43].

The evolution from pristine to composite samples can be divided into two distinct regimes. The first regime corresponds to the transition from $S_1$ (pure $MgB_2$) to $S_2$ ($MgB_2$ with non-emissive GaP particles). In this stage, the $E_{2g}$ peak undergoes a slight blue shift (hardening) accompanied by a minor linewidth adjustment, and the corresponding electron-phonon coupling constant $\lambda$ decreases slightly. This indicates that when GaP acts solely as a "non-emissive inhomogeneous phase," its primary effect is to introduce local strain and interfacial scattering, producing modest perturbations in the phonon potential landscape. These perturbations slightly increase the effective force constant of the $E_{2g}$ mode and consequently weaken the electron-phonon coupling. This interpretation is consistent with transport observations: $\rho$(300 K) increases whereas RRR remains nearly unchanged, reinforcing the view that non-emissive inhomogeneous phases act as "interfacial perturbations" rather than "pairing enhancers," a behavior commonly reported in the literature[15, 39, 41-44].

The second regime occurs for composite samples with activated GaP electroluminescence ($S_3$-$S_8$). Upon increasing the electroluminescence intensity, the Raman $E_{2g}$ peak exhibits a systematic red shift and significant broadening. While the pristine MgB$_2$ sample shows an $E_{2g}$ peak at 593 cm$^{-1}$, the peak position progressively shifts to 570 cm$^{-1}$ for the emissive composites, and the FWHM increases from 154 cm$^{-1}$ to over 210 cm$^{-1}$. The red shift indicates softening of the B-B bond-stretching vibrations, whereas the broadening reflects increased phonon-scattering channels and reduced phonon lifetimes. In this regime, GaP no longer serves as a "static nanoscale inclusion"; instead, under an applied bias, it becomes an electroluminescent center generating surface electromagnetic polarization modes and localized near-field evanescent waves at the GaP/MgB$_2$ interface. These near fields couple synergistically with the $E_{2g}$ phonon mode of the Mg-B layers, modifying the phonon potential landscape and enhancing the effective electron-phonon interaction strength. The vibrational signatures, namely the softening of the $E_{2g}$ mode and the broadening of its linewidth, are direct manifestations of this interfacial coupling between the light field and the phonon system[43].

Changes in the low- and high-frequency components further corroborate this coupling mechanism. The low-frequency peak becomes slightly more intense and shifts to higher wavenumbers with increasing GaP content, indicating strengthened interactions between interfacial lattice vibrations and grain-boundary acoustic modes. This behavior is likely associated with strain modulation and variations in the local density of states induced by the GaP nanoparticles. The high-frequency shoulder becomes more pronounced, suggesting enhanced contributions from second-order scattering and interface-related vibrational modes. This indicates the formation of active electromagnetic coupling zones at the GaP/MgB$_2$ interfaces under applied electric fields, strengthening the interaction between high-energy vibrational states and electronic excitations.

To quantify the impact of this spectral evolution on superconducting pairing, the characteristic frequencies obtained from the three-peak fits were used to extract the logarithmic average phonon frequency. The electron-phonon coupling constant $\lambda$ was then calculated using the Allen-Dynes method as described in the Experimental Section[45]. The results show that $\lambda$ decreases slightly from S$_1$ to the non-emissive S$_2$, consistent with the modest $E_{2g}$ hardening. However, from S$_3$ onward, $\lambda$ increases systematically from 0.856 to 0.889 as the electroluminescence intensity rises, tightly correlating with the observed enhancement in $T_c$. In other words, introducing GaP as a mere "static inhomogeneous phase" only introduces interfacial strain and scattering, leading to a slight reduction in $\lambda$; only when GaP is electrically excited—producing sustained interfacial optical fields and near-field electromagnetic modes—does $\lambda$ increase beyond that of pristine $MgB_2$, driving the monotonic rise in $T_c$.

Collectively, the Raman spectra reveal two fundamentally different physical roles of inhomogeneous phases in $MgB_2$: non-emissive inhomogeneous phases primarily introduce static strain and interfacial scattering that mildly suppress pairing, whereas electroluminescent inhomogeneous phases enhance electron-phonon coupling through interfacial light-field-phonon cooperation under bias excitation. The sequence that includes the blue shift of the $E_{2g}$ mode in sample $S_2$, followed by the red shift and linewidth broadening in the emissive samples, together with the initial decrease and subsequent increase in $\lambda$ that parallels the evolution of Tc, presents a coherent physical picture. The GaP electroluminescent inhomogeneous phases activate the key phonon modes of $MgB_2$ through near-field coupling and mitigate the negative effects typically associated with conventional inert nanoparticle additions, ultimately strengthening the superconducting pairing channel[16].

It should be further clarified that the energy of visible photons (~eV) is much larger than the energy scales of the superconducting gap and phonons (~meV). However, the mechanism proposed in this work is not that "photons directly supply pairing energy."

Instead, the electroluminescence at the interface generates optical/electromagnetic near fields on the nanoscale, which modify the lattice potential energy surface and the phonon self-energy, thereby renormalizing the $E_{2g}$ mode and enhancing the electron-phonon coupling constant $\lambda$. Within the Eliashberg/Allen-Dynes framework, $T_c$ is jointly determined by $\lambda$ and the characteristic phonon frequency $\omega_{log}$. When the near field induces $E_{2g}$ phonon softening accompanied by a redistribution of spectral weight, $\lambda$ may increase and consequently elevate $T_c$ without significantly enhancing impurity scattering. In the present study, under the condition of a fixed GaP addition level and only minor variations in the XRD Rietveld refinement parameters (Table 1), a monotonic correlation between the $E_{2g}$ frequency, FWHM, and $T_c$ with increasing EL intensity is observed (Figs. 4 and 5). This experimental evidence supports a near-field-driven dynamical modulation mechanism rather than a purely static structural effect.

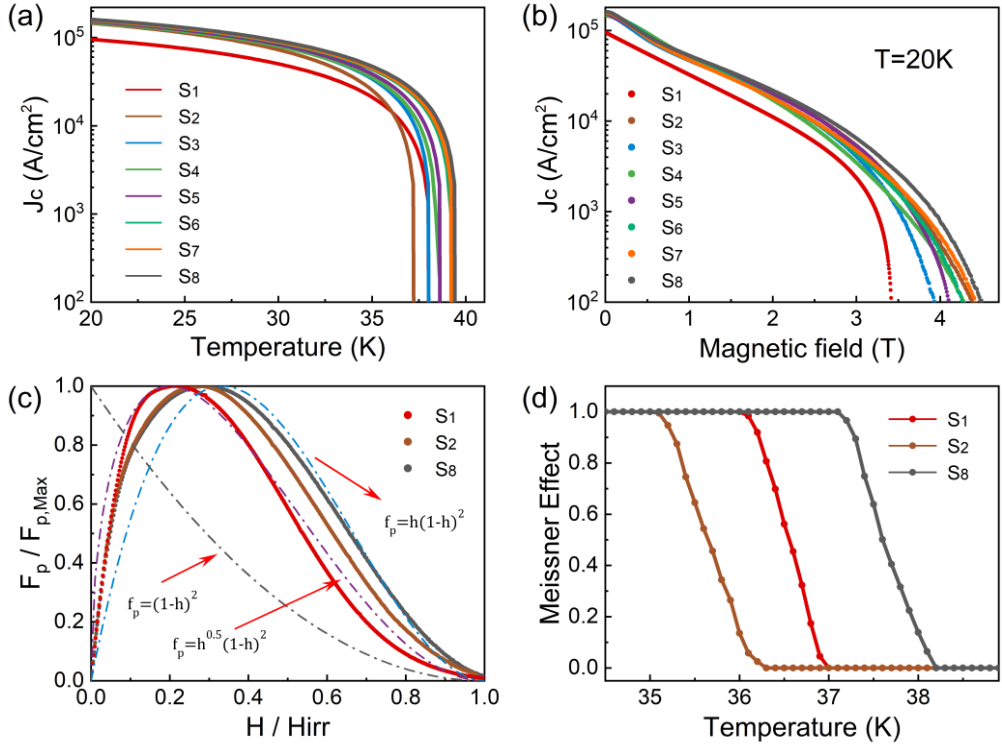

**Figure 6.**
**(a)** $J_c$-$T$ curves of pristine $MgB_2$ and $MgB_2$ composites containing 0.5 wt.% GaP electroluminescent inhomogeneous phases. The composites exhibit higher $J_c$ across the entire temperature range.
**(b)** $J_c$-$H$ curves showing that the composites maintain significantly stronger current-carrying capability in the high-field region.
**(c)** $F_p$-$H$ pinning-force curves. Introduction of GaP inhomogeneous phases increases $F_{p,max}$ and shifts the peak position $h_m$ from 0.218 to 0.272, indicating enhanced pinning strength within the dominant point/grain-boundary pinning regime.
**(d)** Meissner-effect curves illustrating stronger diamagnetic response and magnetic flux expulsion for the composite sample. Static field-repulsion strength estimated from levitation height.

Figure 6 presents the critical current density ($J_c$), pinning force ($F_p$), and Meissner effect for the pristine $MgB_2$ sample and the sample incorporating 0.5 wt.% GaP electroluminescent inhomogeneous phases under various temperatures and magnetic fields. The $J_c$-$T$ curves (Figure 6a) show that under self-field conditions, the pristine $MgB_2$ sample exhibits a $J_c$ of $9.55 \times 10^4$ A·cm$^{-2}$ at 20 K, whereas the composite sample reaches approximately $1.61 \times 10^5$ A·cm$^{-2}$, corresponding to a 69% enhancement. With increasing temperature, $J_c$ for both samples decays exponentially; however, the composite sample consistently maintains higher values across the entire temperature range. This indicates that the GaP electroluminescent inhomogeneous phase promotes sintering-induced densification, optimizes grain boundary structures, and introduces additional nanoscale dispersed phases/defects that act as effective pinning centers, thereby improving the current-carrying pathways and enhancing flux pinning capability.

The $J_c$-$H$ curves (Figure 6b) further reveal the pronounced high-field advantages introduced by GaP inhomogeneous phases. The pristine MgB$_2$ sample shows a rapid reduction in $J_c$ with increasing external field, whereas the composite sample exhibits a notably slower decay. At 2 T, the pristine sample shows a $J_c$ of $1.11 \times 10^4$ A·cm$^{-2}$, while the composite reaches $2.18 \times 10^4$ A·cm$^{-2}$ (an increase of 97%). At 3 T, $J_c$ increases from $2.36 \times 10^3$ A·cm$^{-2}$ in the pristine sample to $6.46 \times 10^3$ A·cm$^{-2}$ in the composite, corresponding to an enhancement of 174%. The observed high-field enhancement of $J_c$ indicates that the GaP inhomogeneous phase forms new, nanoscale effective pinning centers within the MgB$_2$ matrix, suppressing flux-line motion and enhancing flux pinning strength. Since the GaP particles are dispersed along grain boundaries and interface regions, the local strain fields and fluctuations in the mean free path they introduce act as randomly distributed point pinning sources, leading to a more gradual decay of the $J_c$-$H$ curves and an upward shift of $H_{irr}$.

The pinning force $F_p$-$H$ curves (Figure 6c) exhibit a typical single-peak feature. For the pure sample, the peak position $h_m \approx 0.218$, close to the "surface pinning" scenario ($h_m \approx 0.2$) in the Dew-Hughes model, consistent with its small grain size and abundant grain boundaries. Upon GaP addition, $h_m$ shifts to $\approx 0.272$, accompanied by a significant increase in $F_{p,max}$, indicating that point-like pinning contributions introduced by the nanoscale inhomogeneous phase/defects effectively enhance the overall pinning, evolving the system from predominantly grain-boundary pinning to a mixed "grain-boundary + point pinning" regime. It is noteworthy that $h_m$ values in the range 0.2-0.3 generally indicate mixed pinning rather than a single mechanism: $h_m$ closer to 0.2 corresponds to surface-dominated pinning, while values approaching 0.33 reflect stronger point-like ($\delta l$-type) pinning. Therefore, the shift of $h_m$ from 0.218 to 0.272 can be interpreted as the GaP dispersed phase providing additional nanoscale condensation energy and mean free path fluctuations near grain boundaries, raising the flux depinning barriers and extending the effective pinning field range. This interpretation is consistent with the observed mid-to-high field enhancement in the $J_c$-$H$ curves.

Figure 6d shows the Meissner effect and the temperature-dependent variation in static field expulsion estimated from levitation height measurements[20]. The composite sample exhibits an earlier onset of diamagnetism, consistent with the elevated $T_{c,onset}$ observed in resistivity measurements. Upon cooling, both samples achieve full diamagnetic screening, whereas the composite sample consistently displays stronger magnetic repulsion across the entire temperature range, indicating improved global shielding capability.

Importantly, the enhancement trends reported here differ fundamentally from those obtained through conventional chemical doping of $MgB_2$. Chemical doping typically relies on aliovalent substitution or interstitial incorporation, affecting electron-phonon coupling and vortex pinning by tuning carrier density, chemical pressure, or introducing impurity phases. However, such structural perturbations frequently induce lattice distortion, enhanced interfacial scattering, and grain-boundary degradation, which makes it difficult to achieve simultaneous improvement in $T_c$, $J_c$, and magnetic response. Enhancement of one property is often accompanied by the deterioration of another, which is a well-known limitation of most metal-doping and nanoparticle-addition strategies[49-53]. In contrast, the present study demonstrates that the GaP electroluminescent inhomogeneous phases improve $T_c$, $J_c$, and $H_{irr}$ concurrently as their emission intensity increases. The critical temperature increases by approximately 1.4 K, the self-field $J_c$ improves by nearly 70%, and $H_{irr}$ increases by over 30%. This "three-parameter co-enhancement" originates from an entirely different mechanism: rather than relying on static structural perturbations, the electroluminescent inhomogeneous phases generate interfacial optical fields and electromagnetic polarization modes under bias excitation, which couple to the $E_{2g}$ phonon mode and strengthen electron-phonon interactions without introducing additional scattering or grain-boundary degradation. Therefore, the "concurrent enhancement" observed in this work is not a mere repetition of conventional chemical doping effects; rather, it represents a parallel realization of near-field-activated pairing enhancement and structural pinning optimization within the same material system.

Compared with our previous studies and commonly reported inhomogeneous-phase-enhanced $MgB_2$ strategies [33], this work offers clearer physical boundaries and stronger innovation in terms of design rationale and attribution. First, the inhomogeneous phases introduced here maintain "constant element/composition" chemically, with their primary function being the modulation of the periodic structure of the luminescent phase (optical potential periodicity/microstructural periodicity) rather than altering elemental content. Consequently, compared with prior approaches where variations in secondary-phase composition could induce carrier scattering, impurity-phase formation, or stoichiometric deviations, this strategy largely eliminates confounding effects from compositional differences, allowing changes in superconducting performance to be more directly attributed to the "periodic structure → light-field distribution → near-field coupling" mechanism. Second, owing to this periodic structure optimization, the inhomogeneous phases exhibit higher electroluminescence (EL) intensity and stronger local optical fields, elevating the modulation of superconducting condensate parameters to a higher level. We observe that as the EL intensity increases, the key superconducting parameters exhibit more pronounced and reproducible responses, indicating that this strategy goes beyond conventional "pinning/densification" gains and actively enhances the superconducting state under stronger light-field driving. Additional distinctive features and innovations of this work include: (i) a systematic "intensity-response" mapping, using EL intensity as a continuously tunable parameter, which establishes a quantitative correlation between superconducting performance and light-field strength, providing experimental criteria to separate structural densification/defect pinning from light-field activation; (ii) spectral evidence from Raman and related techniques combined with the $T_c$-EL correlation, forming a chain of evidence linking "periodic structure → near-field enhancement → lattice/electron-phonon coupling response → superconducting parameter improvement"; (iii) a decoupled discussion of "EL-driven $T_c/\lambda$ enhancement" and "microstructural pinning-driven $J_c/H_{irr}$ improvement," preventing all enhancements from being attributed to a single mechanism and thus increasing the testability and generalizability of the conclusions.

In summary, this work provides a pathway distinct from traditional chemical doping or secondary-phase element introduction: by maintaining nearly constant chemical composition while leveraging periodic-structure engineering and high-intensity near-field luminescence, it enables deeper and more controllable modulation of the $MgB_2$ superconducting state.

## 3. Conclusion

This study demonstrates that incorporating GaP electroluminescent inhomogeneous phases into $MgB_2$ and activating their emission under bias current effectively enhances the superconducting performance of the host material. As the emission intensity of GaP increases, the superconducting transition temperature of $MgB_2$ rises from 38.2 K to 39.6 K, accompanied by a simultaneous upward shift of the zero-resistance temperature. The self-field critical current density at 20 K improves by approximately 69%, while enhancements of up to 170% are observed in the high-field range of 2-3 T. The irreversibility field $H_{irr}$ also increases by 31.5%. In contrast to conventional chemical doping, where the simultaneous improvement of $T_c$, $J_c$, and magnetic response is notoriously difficult, the introduction of electroluminescent inhomogeneous phases enables a cooperative enhancement of all three parameters and produces a systematic and monotonic dependence on emission intensity. These results indicate that GaP inhomogeneous phases not only improve microstructural connectivity and flux-pinning landscapes but also substantially strengthen current-carrying capability and magnetic shielding performance. Raman spectroscopy, XRD refinement, and SEM analyses collectively reveal that the incorporation of GaP does not disrupt the $MgB_2$ host lattice. Instead, it introduces mild lattice modulation, more compact grain-boundary structures, and additional nanoscale pinning centers. Under bias excitation, the GaP/$MgB_2$ interfaces generate localized electroluminescence and surface electromagnetic polarization modes, which likely give rise to electromagnetic near fields. These near fields can couple to the $E_{2g}$ phonon mode of the Mg-B layers, enhancing electron-phonon interactions and driving the systematic increase in $T_c$. It should be noted that while interfacial light-field-phonon coupling provides a compelling primary mechanism, the experimental observations do not exclude the participation of other physical effects. Although these possibilities require further experimental or theoretical validation, they offer a broader physical context for understanding light-field-driven superconductivity modulation. In summary, this work establishes a new pathway for achieving synergistic enhancement of superconducting properties in MgB$_2$ while preserving its intrinsic lattice framework. The findings highlight the potential of electroluminescent inhomogeneous phases to modulate electron-phonon coupling, vortex dynamics, and key superconducting parameters, providing a conceptual foundation for designing next-generation superconducting materials based on internal light-field regulation. Future work will proceed along two directions: (i) further optimization of the multiquantum-well periodic structures and core-shell parameters to expand the EL intensity window and establish broader dynamic-range calibration curves for $T_c$-EL and $\lambda$-EL correlations; (ii) systematic examination of potential nonlinearities and saturation effects in the superconducting response under high EL intensity, including phenomena such as interface light-field saturation/secondary absorption, carrier recombination limitations, and Joule heating, thereby refining a predictive "intensity-response" theoretical framework.

## 4. Experimental Section
### 4.1 Preparation of GaP Electroluminescent Particles

GaP electroluminescent inhomogeneous phases were synthesized via a hot-injection method. The precursor system was prepared by dissolving gallium acetylacetonate, tris(trimethylsilyl)phosphine, zinc acetylacetonate, tellurium powder, and indium chloride in octadecene (ODE) according to predetermined molar ratios. The reaction was carried out in a three-neck flask under an inert atmosphere. After heating the mixture to 300 °C, tris(trimethylsilyl)phosphine and trioctylphosphine (TOP) were swiftly injected to initiate nucleation and the subsequent growth of multilayer core-shell structures. By controlling the injection sequence and the dwelling time, two classes of multilayer core-shell units, namely $[GaP:Zn/GaP]_{\times m}/[GaInP/GaP]_{\times n}$ and $[GaP:Te/GaP]_{\times m}/[GaInP/GaP]_{\times n}$, were constructed, and these units eventually formed nanoparticles with diameters of approximately 7 nm during the later stage of growth.

Upon completion, the reaction mixture was naturally cooled to room temperature and washed several times with ethanol via centrifugation to remove unreacted precursors. The resulting wet product was dispersed and dried to form a gel-like precursor film, which was subsequently subjected to high-temperature treatment in a tubular furnace under flowing Ar to eliminate organic residues and promote the aggregation of nanocrystals into electroluminescent particles. The final product was ball-milled to narrow the particle-size distribution to 400 nm, yielding GaP electroluminescent inhomogeneous phase particles suitable for composite fabrication.

To obtain GaP inhomogeneous phases with varying intrinsic luminescence intensities, we tuned the number of repetitions of the multilayer core-shell units (m, n) to modify carrier recombination pathways and radiative efficiency, thereby producing a series of particles with different emission efficiencies. After high-temperature treatment and ball milling, each batch of particles was excited under a constant bias current of $I_{bias}$ = 100 mA using a custom-built testing fixture, and the electroluminescence (EL) spectra were collected via a fiber-optic spectrometer. In this work, the "EL intensity" is defined as the integrated area of the emission spectrum over the 500-750 nm range (arbitrary units, a.u.), based on which particles with EL = 0, 1000, 2200, 3960, 4900, 5800, and 6600 were selected for compounding with $MgB_2$.

**4.2 Preparation of $MgB_2$ and $MgB_2$-GaP Composite Samples**

Both the $MgB_2$ matrix and $MgB_2$-GaP composite samples were prepared using a non-in situ sintering process. The selected GaP content (0.5 wt.%) was based on prior optimization experiments. All weighing, mixing, and transfer operations were performed inside an Ar-filled glovebox with $O_2$ and $H_2O$ levels < 0.1 ppm to avoid oxidation of $MgB_2$. Commercial $MgB_2$ powder (99.9%, Alfa Aesar) and the GaP inhomogeneous phase particles were separately dispersed in ethanol and ultrasonicated for 20 min to prevent agglomeration. Each dispersion was then magnetically stirred for 20 min to form stable suspensions. The GaP suspension was subsequently added dropwise into the $MgB_2$ suspension and stirred for an additional 20 min to ensure uniform dispersion. The mixed suspension was dried under vacuum at 60 °C for 4 h to obtain black composite powder, which was ground for 1 h in an agate mortar to further improve homogeneity. The composite powder was cold-pressed into cylindrical pellets (diameter: 11 mm; height: 1.2 mm) under a pressure of 14 MPa and sealed in tantalum containers to prevent contamination during sintering. Sintering was conducted in a tubular furnace under ultra-high-purity Ar: the pellets were heated to 850 °C at a rate of 5 °C min$^{-1}$ and held for 10 min, followed by cooling to 650 °C at −5 °C min$^{-1}$ and annealing for 1 h. The samples were finally furnace-cooled to room temperature. All samples contained a fixed GaP loading (0.5 wt.%) and were labeled $S_1$-$S_8$ according to the electroluminescence intensity of the GaP inhomogeneous phases used, enabling systematic investigation of the influence of luminescence strength on the superconducting properties of $MgB_2$.

### 4.3 Characterization Methods

The crystal structure of the samples was analyzed using an X-ray diffractometer (Bruker D8 Advance, Cu Kα radiation, $\lambda$ = 1.5418 Å). Microstructural morphology was examined by scanning electron microscopy (SEM, FEI Verios G4), and elemental distribution was obtained via energy-dispersive X-ray spectroscopy (EDS, Thermo NS7). Electrical transport properties were measured using a standard four-probe method, with low-temperature conditions provided by a closed-cycle cryogenic system (Advanced Research Systems, base temperature ≈ 10 K). Magnetic measurements were performed on a physical property measurement system (Cryogenic CFMS-14T). The magnetic critical current density $J_c$ was calculated from the magnetic hysteresis loops (M-H) using the Bean model:

$$J_c = \frac{20\Delta M}{a^2 bh(1 - \frac{a}{3b})} \ (with\ b > a)$$

where $\Delta M$ is the width of the *M-H* loop (emu), and a, b, and h represent the width, length, and thickness of the sample, respectively. Samples used for magnetic measurements were cut from the central region of the cylindrical pellets, with typical dimensions of 1.5 mm × 2 mm × 1 mm. During measurement, the ab-plane was oriented perpendicular to the external magnetic field, and the field sweep rate was set to 50 mT s$^{-1}$. In this study, the irreversibility field $H_{irr}$ was defined as the magnetic field at which $J_c$ drops to $10^2$ A·cm$^{-2}$. Raman spectra were collected on a WITec Alpha300R confocal micro-Raman system using a TEM$_{00}$ 532 nm laser as the excitation source and a 600 g mm$^{-1}$ grating for spectral acquisition.

## 4.4 Calculation of Electron-Phonon Coupling Constant ($\lambda$)

The electron-phonon coupling constant $\lambda$ was determined from the characteristic phonon frequencies obtained through three-peak Gaussian fitting of the Raman spectra. Raman spectra in the 200-1000 cm$^{-1}$ range were fitted with three Gaussian components corresponding to the low-frequency acoustic mode, the mid-frequency $E_{2g}$ mode, and the high-frequency second-order/interface-related vibrational mode. The peak positions $\omega_i$ and peak areas $A_i$ extracted from the fitting were used to calculate the weight of each phonon mode in the overall coupling process:

$$w_i = \frac{A_i}{\sum_j A_j}$$

The logarithmic averaged phonon frequency was then obtained according to the Allen-Dynes approximation:

$$\langle w_{log} \rangle = exp\left(\sum_i w_i ln w_i\right)$$

The relationship between the superconducting transition temperature $T_c$ and the electron-phonon coupling constant $\lambda$ was determined using the Allen-Dynes modified McMillan formula:

$$T_c = \frac{\langle \omega_{log} \rangle}{1.2} exp\left(-\frac{1.04(1+\lambda)}{\lambda - \mu^*(1+0.62\lambda)}\right)$$

where $\mu^* = 0.13$ is the commonly used Coulomb pseudopotential. By substituting the experimentally measured $T_c$ and $\langle \omega_{log} \rangle$ into the above equation, $\lambda$ can be obtained. This approach enables the softening and broadening of the $E_{2g}$ phonon mode observed in Raman spectra to be quantitatively translated into the electron-phonon coupling strength, providing a reliable basis for analyzing the influence of interfacial light-field modulation on superconducting pairing.

## Acknowledgements

This research was supported by the National Natural Science Foundation of China for Distinguished Young Scholar under Grant No. 50025207 and the National Natural Science Foundation of China Grant No. 52272306.

## Conflict of Interest

The authors declare no conflict of interest.

## Data Availability Statement

The data that support the findings of this study are available from the corresponding author upon reasonable request.